# Lattice dynamics and Raman spectrum of supertetragonal PbVO$_3$.

P. Bouvier[1*], A. Sasani[2], E. Bousquet[2], M. Guennou[3], J. Agostinho Moreira[4]

[1] *Université Grenoble Alpes, Institut Néel CNRS, 25 Rue des Martyrs, 38042, Grenoble, France.*
* pierre.bouvier@neel.cnrs.fr

[2] *Université de Liège, CESAM, QMAT, Physique Théorique des Matériaux, 19 Allée du 6 août, B-4000, Sart Tilman, Belgium.*

[3] *Department of Physics and Materials Science, University of Luxembourg, 41 rue du Brill, L-4422 Belvaux, Luxembourg.*

[4] *IFIMUP – Instituto de Física de Materiais Avançados, Nanotecnologia e Fotónica, Departamento de Física e Astronomia, Faculdade de Ciências da Universidade do Porto. Rua do Campo Alegre s/n. 4169-007 Porto, Portugal.*

Abstract:

Lead vanadate PbVO$_3$ is a polar crystal with a *P4mm* space group at ambient conditions. It is isostructural with the model soft-mode driven ferroelectric PbTiO$_3$, but differs from it by the so-called "supertetragonal" elongation of its unit cell. In this paper, we report a combined study of the lattice dynamics of PbVO$_3$ by Raman spectroscopy at room temperature and first-principle calculations. All zone-center transverse optical (TO) phonon modes are identified by polarized, angle-dependent Raman spectroscopy and assigned as follows: E modes at 136, 269, 374 and 508 cm$^{-1}$, A$_1$ modes at 188, 429 and 874 cm$^{-1}$ and B$_1$ mode at 319 cm$^{-1}$. The calculations confirm the experimental symmetry assignment and allow to obtain the longitudinal (LO) phonons wavenumbers. Besides, we analyze the mode eigenvectors in detail, in order to identify the atomic displacements associated with each mode and compare them with PbTiO$_3$. In spite of their differences in chemistry and strain, the phonon eigenvectors are found to be remarkably comparable in both compounds. We discuss the position of the ferroelectric soft mode in PbVO$_3$ as compared to PbTiO$_3$. A sizeable splitting of the B$_1$+E modes appears as a characteristic feature of supertetragonal phases. The peculiarity of the vanadyl V-O bond frequency in PbVO$_3$ is also addressed.

## 1. Introduction

Polar tetragonal perovskites have been attracting the attention of the scientific community because of their high electric polarization and piezoelectric coefficients which make them promising for functional devices. The model ferroelectric PbTiO$_3$ crystallizes in the *P4mm* space group at room conditions; it can reach an electric polarization as large as 57(3) to 70 µC.cm$^{-2}$, respectively measured or calculated [1,2], due to the combination of both Ti$^{4+}$ cations and stereochemical active Pb$^{2+}$ 6s$^2$ lone pair electrons off-centering displacements. Besides, the electric polarization is coupled cooperatively to the tetragonal strain [2], which gives a highly elongated unit cell that strongly enhance the polarization. This effect has attracted attention on the family of so-called "supertetragonal" polar perovskites that include PbVO$_3$, BiFeO$_3$ films under compressive strain [3], BiCoO$_3$ [4], Bi$_2$ZnVO$_6$ [5], solid solutions between them [6,7] or even BiMnO$_3$ between 40 and 55 GPa [8]. They are characterized by an elongation ratio c/a around 1.2, way larger than in classical ferroelectric perovskites where it reaches at most 1.08. Besides, many of these compounds – PbVO$_3$ included – are also magnetic, which makes them potentially attractive multiferroics.

PbVO$_3$ can be synthesized at high temperatures and high pressures, and recovered at room conditions in the tetragonal *P4mm* structure with a= 3.80005(6) Å, c= 4.6703(1) Å and c/a ≈ 1.23 [9,10]. The PbVO$_3$ structure is similar to the perovskite. However, due to the large tetragonal deformation and the presence of a short V-O vanadyl bond length, the three-dimensional octahedral network is rather considered as layers of corner-shared VO$_5$ pyramids [9,10]. The polar structure is stable from 12 to 570 K, above which it oxidizes to Chervetite Pb$_2$V$_2$O$_7$ [9,10]. The polar nature was also confirmed by second-harmonic generation measurements in the 4-400 K temperature range [11]. Upon increasing hydrostatic pressure, PbVO$_3$ remains stable up to 2.5 GPa before collapsing into a cubic *Pm-3m* perovskite structure via a first-order transition with a volume drop of 10 % [10,12]. The exceptionally large tetragonality of PbVO$_3$ suggests that its ferroelectric polarization should be very high; a value as high as 150 $\mu$C/cm$^2$ was predicted from Berry-phase calculations [13]. To the best of our knowledge, ferroelectric switching has not yet been demonstrated in PbVO$_3$, but past experience has shown that switching of supertetragonal BiFeO$_3$ films was possible [14] and we anticipate that PbVO$_3$ could be switched as well in suitable conditions.

With its V$^{4+}$ cation in a *3d$^1$* configuration, PbVO$_3$ is expected to be magnetic but there is still no consensus about its magnetic structure. First-principle calculations have predicted that PbVO$_3$ should exhibit a two-dimensional antiferromagnetic ground state (of C-type or G-type) as a result of the ordering of the d$_{xy}$ electronic orbitals [13,15,16]. Experimentally, an antiferromagnetic ordering has been proposed from magnetization measurements and muon spin rotation in the 43-50 K range [16]. On the other hand, magnetic susceptibility and specific heat measurements have suggested that PbVO$_3$ exhibits a magnetic frustration and should rather be considered as a disordered ½ spin square lattice [17]. No sign of ferromagnetic ordering could be observed [18] nor calculated [19] under high-pressure.

The lattice dynamics of PbVO$_3$ have been scarcely studied and little is known about its Raman signature. According to Singh's calculations, PbVO$_3$ has three fully symmetric modes at 190, 408 and 838 cm$^{-1}$ in the *P4mm* structure; the last and highest frequency mode being assigned to the short and stiff V-O vanadyl bond vibration [15]. An unpolarized Raman spectrum of a PbVO$_3$ powder was published by Okos *et al.* [20]. They observed at least eight modes at 187, 210, 263, 316, 370, 506, 876 and 942 cm$^{-1}$ and have proposed a symmetry assignment. The Raman spectra of PbVO$_3$ thin films deposited on (001)-oriented LaAlO$_3$ substrates display two broad bands at 830 cm$^{-1}$ and 930 cm$^{-1}$ assigned to the TO-LO components of the short V-O symmetrical stretching mode, and an intense band at 593 cm$^{-1}$ assigned to the long V-O stretching mode [21]. Additional peaks at low wavenumbers were reported at 390, 350, 220 and 190 cm$^{-1}$ and assigned to V-O-V bending modes and vibrations involving the heavy Pb cation [21]. Similarly, broad bands with weak intensity were reported on other PbVO$_3$ thin films [22]. Besides, those measurements show the presence of a double peak at 965 cm$^{-1}$ which is usually associated with other vanadate species (VO$_x$). Overall, the lack of a comprehensive assignment of the Raman-active phonons in PbVO$_3$ hinders the use of Raman spectroscopy for investigations of the physics of PbVO$_3$.

In this paper, we address this problem by combining an angle-resolved, polarized Raman spectroscopy study on a PbVO$_3$ single crystal at room temperature with ab initio calculations of its lattice dynamics. We show how this approach allows us to successfully identify all phonon modes and their symmetries. Besides, we address the relevance of the comparison between lattice dynamics of PbTiO$_3$ and PbVO$_3$ by a detailed analysis of the calculated eigenvectors.

2. Methodological details

The crystal used in the Raman experiments originates from the growth batch described by Okos et al. [20] using a two-stage solid-state reaction process at high-temperature (950°C) and high-pressure (6 GPa) in a CONAC type apparatus. We have performed preliminary characterization of several crystals from this batch by single crystal x-ray diffraction with a D8Venture diffractometer (λ=0.71073 Å). The Bragg peaks were indexed using the APEX4 v2022 software from Bruker™. From one crystal to another, the quality of the pattern did not always allow for a full refinement of atomic positions, because of twinning or other issues, but the unit cell a=b=3.8043(15) Å and c=4.6773(23) Å and crystal symmetry were always consistent with previous structural refinements [9,10,20].

The Raman spectra were recorded at room conditions in the backscattering geometry, using the final stage of a T64000 Jobin-Yvon spectrometer equipped with 1800 /mm grating blazed at 500 nm. The 514.5 nm and 488 nm lines of an $Ar^+$ laser (Spectra Physics Stabilite2017) were chosen for excitation. Semrock™ Maxline and Razor edge filters were used before entering the spectrometer to clean and reject the stray laser line with a cut-off wavenumber of 80 $cm^{-1}$. The laser was focused on the sample surface with an x50 objective (Olympus NA=0.75), which allows a 2 μm diameter laser spot on the sample. During our preliminary studies, we observed that $PbVO_3$ can transform into Chervetite ($Pb_2V_2O_7$) under intense laser irradiation. The Chervetite exhibits a totally different Raman spectrum, easily recognized by two intense bands centered at 816 and 876 $cm^{-1}$ (see Figure S.1 in Supplemental Information). The decomposition of $PbVO_3$ in Chervetite is expected at a temperature higher than 570 K [9,10], which can definitively be reached by the laser as the compound is not transparent. We found that reducing the incident laser power on the sample as low as 0.4 mW (as measured after the objective), using a broadband neutral density filter, avoids sample degradation.

The crystal was centered in a specific holder allowing to rotate the sample over 360° around the vertical axis, coincident with the incident and scattered light directions, ensuring that measurements are performed at the exact same position on the sample surface during the entire 360° rotation.

For the polarized Raman measurements, we consider two reference frames shown in Figure 1: a laboratory reference frame (*X*,*Y*,*Z*) and a sample reference frame (*x*,*y*,*z*) with x//a-axis, y//b-axis and z//c-axis. It will be shown later in the results section that the sample is oriented with a crystallographic *b*-axis along *Y,* so that both frames share a common vertical *Y*=y axis that also corresponds to the microscope optical axis. In the laboratory reference frame (*X*,*Y*,*Z*), the laser polarization directions labelled V (vertical) and H (horizontal) are along *X* and *Z*, respectively. The crystallographic orientation of the sample in plane is initially unknown, and the angle between the *Z*- and *z*-axes is defined as $\theta_o$. This value is found *a posteriori* from the analysis of the angular dependence of the Raman intensity measured under fixed polarizations conditions while rotating the sample over an azimuthal angle $\theta$ from 0° to 180°. The angle dependence was measured in all three y(VV)y, y(HV)y and y(VH)y configurations, in the Porto notation. Using IgorPro® software, a sum of independent damped oscillators multiplied by the Bose-Einstein factor [23] was fitted to the Raman spectra to extract the angular dependence of the wavenumber, the half width at half maximum and the intensity (integrated area) of each individual band. One band displayed a significantly asymmetric shape and was fitted instead by a Fano line shape, as discussed separately below. The integrated area was then calculated by integrating the area below the curve.

Our first-principles density functional theory (DFT) study of $PbVO_3$ was done using the CRYSTAL17 software package [24]. We have used the HSE06 hybrid functional for the exchange correlation term [25,26]. The structural parameters and phonon frequencies were found to converge by using a 8x8x8 *k*-points mesh sampling and the structure was relaxed until the forces are less than $3.10^{-5}$ Ha/Bohr. The phonon calculations were done using the frozen phonon technique as implemented in Crystal17 [27,28]. To calculate the non-analytical correction to the phonon frequencies along the [001] and [100]/[010] propagation directions (LO-TO splitting) we have used the Abipy and the Anaddb postprocessing scripts of the ABINIT packages [29]. We have used the Gaussian type basis sets from reference [30], [31] and [32] for Pb, V and O, respectively, which give the best structural agreement with experiment. The truncation thresholds in the evaluation of the Coulomb and exchange series present in the CRYSTAL code were adjusted to $10^{-7}$ for Coulomb overlap tolerance, $10^{-7}$ for Coulomb penetration tolerance, $10^{-7}$ for exchange overlap tolerance, $10^{-7}$ for exchange pseudo-overlap in the direct space, and $10^{-14}$ for exchange pseudo-overlap in the reciprocal space. The symmetry of the different structures and the phonon modes were determined using the FINDSYM software [33]. The calculation of the energy of the different magnetic phases shows that the C-type antiferromagnetic ordering is the ground state in agreement with previous DFT calculations [9,13,15,16,17], such that all the calculations reported below were performed in this magnetic phase unless stated otherwise.

3. Results and discussion

3.1. Group theory prediction of Raman modes symmetry

Factor group analysis of the structure (4*mm*, $C_{4v}$ point group) yields the following decomposition of the optical normal modes in eight irreducible representations at the Γ-point of the Brillouin zone:

$$\Gamma^{opt} = 3A_1 + B_1 + 4E \tag{1}$$

The Wyckoff positions occupied by the different atoms in the primitive cell and their contributions to the Raman-active modes are given in Table 1. The $A_1$ and E modes are simultaneously infrared and Raman-active, while the $B_1$ mode is only Raman-active. Because the $A_1$ and E modes are infrared-active, the transverse (TO) and longitudinal (LO) optical modes split according to the Lyddane-Sachs-Teller relation [34,35]. The Raman selection rules are identical to the $PbTiO_3$ case [36,37,38] and are recalled briefly here. The Raman tensors *T* in the 4*mm* ($C_{4v}$) point group can be written as follows:

$$A_1(z) = \begin{pmatrix} \alpha & 0 & 0 \\ 0 & \alpha & 0 \\ 0 & 0 & \beta \end{pmatrix} \tag{2}$$

$$B_1 = \begin{pmatrix} \gamma & 0 & 0 \\ 0 & -\gamma & 0 \\ 0 & 0 & 0 \end{pmatrix} \tag{3}$$

$$E(x) = \begin{pmatrix} 0 & 0 & \delta \\ 0 & 0 & 0 \\ \delta & 0 & 0 \end{pmatrix} \tag{4}$$

$$E(y) = \begin{pmatrix} 0 & 0 & 0 \\ 0 & 0 & \delta \\ 0 & \delta & 0 \end{pmatrix} \tag{5}$$

where α, β, γ and δ are considered as real tensor elements. From them we can derive the modes allowed and Raman intensities expected in relevant backscattering configurations, as summarized in Table 2. With a laser propagation direction along the tetragonal c-axis, 4 modes ($3A_1(LO)+B_1$) are expected while up to 8 modes can be expected if the laser propagation is perpendicular to the c-axis. Importantly, E(LO) modes are not accessible in a backscattering configuration. This is at variance with uniaxial 3*m* polar crystals like $LiNbO_3$ where the rhombohedral symmetry makes E(LO) modes accessible in backscattering [39]. Measuring those modes would require other geometries like right angle scattering or platelet geometry [38,40] that are very difficult to achieve here due to sample size and the strong absorption of the crystals.

It is also useful to recall that the $A_1$ and E modes originate from the three optical polar modes of the $T_{1u}$ irreducible representation in the $O_h$ point group of the parent cubic perovskite (space group *Pm-3m*, Z=1), whereas the $B_1$ and the fourth E modes originate from the silent $T_{2u}$ irreducible representation [36,37]. In the following, we will therefore label the modes when necessary like it was done in past studies of PbTiO$_3$ [36,37,38]: the modes originating from the $T_{1u}$ modes will be indexed with a number 1, 2 and 3 in order of increasing frequency, for example $A_1$(1TO), $A_1$(2TO) and $A_1$(3TO).

3.2. Raman spectra of PbVO$_3$ at room conditions and sample orientation

Several crystals were measured and a crystal was selected that showed only weak signal from the spurious Chervetite phase and a minimal number of well-defined sharp bands. Representative Raman spectra of PbVO$_3$ recorded in the HV-scattering geometry using the 514.5 nm and the 488 nm laser lines as excitation, are shown in Figure 2. Both spectra display eight well-defined bands which matches the expectations from the group theory analysis for a laser propagation direction perpendicular to the tetragonal c-axis. The 514.5 nm excitation line was selected for all further experiments. It is worth stressing that the low wavenumber band at 136 cm$^{-1}$ is sometimes hidden by the scattered signal arising from rotational modes of O$_2$/N$_2$ molecules present in the air. The observed very low intensity broad bands between 750 and 900 cm$^{-1}$ are associated to small traces of spurious phase, likely associated with Chervetite secondary phase formed during the synthesis process.

3.3. Angular dependence of the Raman intensity and symmetry assignment.

In order to confirm the crystal orientation and make a symmetry assignment of the Raman modes, we performed angle-dependent Raman measurements by rotating the crystal along the optical axis of the microscope in both parallel and crossed polarization conditions. Representative Raman spectra recorded in the 100 - 1000 cm$^{-1}$ spectral range in the VV- and HV-scattering geometries for several selected azimuthal angles, are presented in Figure 3. The intensity of the Raman bands depends on both the scattering geometry and the orientation of the incident electric field relatively to the *a*- and *c*- crystallographic axes, defined by the azimuthal angle $\theta + \theta_o$ (see Figure 1). The angular dependence of the Raman intensity is given by the following equation:

$$I(\theta) \propto |\hat{e}_i \mathcal{R}(\theta).T.\mathcal{R}^{-1}(\theta)\hat{e}_s|^2 \qquad (6)$$

where $\hat{e}_i$ and $\hat{e}_s$ are the unit vectors of the incident and scattered electric field, respectively, and $\mathcal{R}(\theta)$ is the rotation matrix referred to the laboratory reference frame. $T$ is the Raman tensor written in the crystal principal axes.

Assuming the experimental geometry shown in Figure 1, the rotation matrix about the *Y*-axis is:

$$\mathcal{R}(\theta) = \begin{pmatrix} \cos(\theta + \theta_o) & 0 & \sin(\theta + \theta_o) \\ 0 & 1 & 0 \\ -\sin(\theta + \theta_o) & 0 & \cos(\theta + \theta_o) \end{pmatrix} \qquad (7)$$

where $\theta_o$ accounts for the initial angle between the z-axis (// to *c*-axis) and the Z-axis (// to H-axis) (see Figure 1). Using equations (2) to (7), we calculated the angular dependence of the Raman intensities for all symmetry modes in each scattering geometries, as shown in Table 3.

According to Equations (8) to (16), the $A_1$, $B_1$ and E modes can be clearly distinguished from the angular dependence of the spectral intensity in both the VV and HH configurations. For instance, each of the $A_1$ and the $B_1$ modes displays twofold symmetric lobes, whereas the E modes display fourfold symmetric lobes. The $A_1$ and $B_1$ modes can also be distinguished as their lobes are rotated by 90º relatively to each other. The E modes are distinguished from the $A_1$ and $B_1$ modes as their lobes are rotated by 45º from the $A_1$ or $B_1$ modes. In cross-polarization geometry, HV (or VH), the intensity of both $A_1$ and $B_1$ modes exhibits the same fourfold symmetric lobes, so that they cannot be distinguished. These considerations will be used to assign a symmetry to the observed Raman bands of $PbVO_3$.

### 3.3.1. E-modes

Figure 4 shows the polar plot of the angular dependence of the intensity of the Raman bands centered at 136, 269, 374 and 508 cm$^{-1}$, measured in the VV- and VH-scattering geometries, respectively. In all cases, $I(\theta)$ exhibits fourfold symmetric lobes, whereby the VH lobes are rotated by 45° with respect to the VV lobes. The solid lines were obtained by a fit of equations (14-16) to the experimental results. The excellent agreement between experiment and theory allows us to unequivocally assign an E symmetry to these modes. The same initial position $\theta_o$ = 33(1)º is determined for all the four E modes in all considered geometries. The $\delta$ values of the Raman tensor, extracted for the four E modes, are reported in Table S.2 in Supplemental Information. These values are dependent on the scattering geometries and other experimental parameters. No attempt was done to correct for this instrument response.

### 3.3.2. $A_1$ modes

Figure 5 shows the polar plot of the angular dependence of the intensity of the Raman bands at 188, 429 and 874 cm$^{-1}$, measured in the VV and VH-scattering geometries, respectively. In the VV-scattering geometry, the intensity of the bands at 188 and 874 cm$^{-1}$ displays two prominent broad lobes, whereas for the band at 429 cm$^{-1}$ it exhibits a fourfold lobe structure. The orientation of the four lobes is rotated by 45º relatively to the lobes observed for the E modes, which excludes the possibility that this band has E symmetry. We notice that the width and shape of the lobes for the $A_1$ modes are strongly dependent on the values taken by α and β tensor coefficients. If these values are close to each other, the twofold lobe will become a circle, showing a $I(\theta)$ independent on the azimuthal angle. However, if α ≈ - β, we expect lobes with four-fold symmetry. This is exactly what we observed for the three modes, allowing us to assign these bands to $A_1$ modes. The solid lines in Figure 5 were determined by the best fit of equations (8-10) to the experimental results. The same initial angle $\theta_o$ = 35(3)º is found for all the three $A_1$ modes in all considered geometries. The α and β element values of each $A_1$ Raman tensor are reported in Table S.2 in Supplemental Information. Notice that the ratio between the tensor elements is β/α=1.19(2) for the 188 cm$^{-1}$ mode, β/α=1.81(5) for the 874 cm$^{-1}$ mode, and β/α= - 0.91(4) for the 429 cm$^{-1}$ band, with one tensor element taking a negative value. Interestingly, the anisotropy of the Raman tensor at 874 cm$^{-1}$ is much larger than for the two other modes. In the cross-polarization VH (or HV) geometries, the fourfold lobe of the 188 and 874 cm$^{-1}$ modes are not perfectly reproduced by equation (8) that should produce four symmetric lobes. This cannot point to a wrong symmetry assignment – all symmetries should lead to four symmetric loves, as can be seen from Table 3 – but rather to an experimental issue in the measurement of intensities. The precise reason for this is not certain at this stage, but it does not compromise our general mode assignment.

3.3.3. $B_1$ mode

The band at 319 cm$^{-1}$ has an asymmetric shape resembling the Fano-type interference [41]. We used this Fano asymmetric function to successfully fit this mode as displayed in Figure S.3 in Supplemental Information. The polar plot of the angular dependence of the intensity of this Raman band, measured in the VV and VH-scattering geometries, is presented in Figure 6. The lobes of largest amplitude in the VV-geometries are perfectly described by equation (11) and are oriented along a direction that is rotated by 90º relatively to the $A_1$ modes, as it is expected for $B_1$ modes. The same initial angle of $\theta_o$ = 34(2)º is found from the position of the maximum intensity of the lobes. The γ values of the Raman tensor are reported in Table S.2 in Supplemental Information. Like before, we notice that the fourfold lobes are not perfectly reproduced by the equation (12) in crossed polarization. Despite this discrepancy, we have assigned a $B_1$ symmetry to this mode.

The asymmetric shape of this $B_1$ mode is reminiscent of other cases in related compounds. Similarly asymmetric shapes were also reported for the $A_1$ TO phonon in polar *P4mm* BaTiO$_3$ at 170 cm$^{-1}$ [42], or at 147 cm$^{-1}$ in PbTiO$_3$ [38,43] or for the 2TO phonon in the non-polar *I4/mcm* SrTiO$_3$ at low temperature [44], with debated explanations about their possible origin. Here, we observe that the peak shape is independent on the excitation wavelength, which exclude any resonant electron-phonon origin. Also, the asymmetrical shape is located at higher energy in the supertetragonal PbVO$_3$ compared to other regular polar *P4mm* perovskites. This reinforces the idea that there may be a coupling with second order acoustical phonons state (2TA) that shift to higher energy as the c/a ratio increase, as it has already been proposed in polar perovskite [42].

3.4. Lattice dynamical calculations for PbVO$_3$ and comparison with experiment

Table 4 summarizes the experimental wavenumber values of the observed TO Raman-active modes and the corresponding symmetry assignment, and compares with the modes wavenumbers obtained from our first-principle calculations. The agreement between experimental and calculated results (for the C-type magnetic ordering) is very good. The schematic representations of the atomic displacements involved in each phonon mode obtained from our DFT simulations are presented in Figure 7. The eigendisplacements are reported in Table S.4 in the Supplemental Information. The largest discrepancy between simulation and experiment is found for the $A_1$(3TO) mode calculated at 910 cm$^{-1}$, 36 cm$^{-1}$ higher than the experimental value (+4%). This $A_1$ mode only involves the vanadyl bond V-O(1) stretching, with a bond length $d_{V-O}$=1.668(8) Å as reported from original X-ray diffraction measurement on PbVO$_3$ [9,10]. According to the Badger's rule $v^{2/3} \propto d^{-1}$ [45], such stretching frequency $v$ of a chemical bond is an inverse function of the bond length $d$ involved in the vibration. Thus the comparison with other vanadyl containing compounds such as CaV$_2$O$_5$ ($\omega$ = 932 cm$^{-1}$; $d_{V-O}$= 1.645 Å) [46], α'-NaV$_2$O$_5$ ($\omega$=972 cm$^{-1}$ ; $d_{V-O}$=1.60 Å) [47], α-V$_2$O$_5$ ($\omega$ = 997 cm$^{-1}$; $d_{V-O}$=1.585 Å) [48,49,50] (see Figure S.5 in Supplemental Information) shows that the V-O bond in PbVO$_3$ is not only longer but also weaker, as compared to other vanadate compounds. In the early calculations by Singh [15], this mode was found at an even lower position of 838 cm$^{-1}$ (see Table 4). The secondnext highest deviation is found for the E(3TO) mode at 491 cm$^{-1}$ with a 17 cm$^{-1}$ downshift with respect to the experimental value. Other modes show deviations smaller than 7 cm$^{-1}$.

The discrepancies between the experimental and calculated wavenumbers can originate from different factors: (i) the exchange correlation functional used here, (ii) the fact that the calculations were done within the C-type antiferromagnetic phase while the Raman experiments were done in the paramagnetic phase (spin-phonon coupling) and (iii) the fact that the calculations were done at 0 K while the experiments were done at room temperature (thermal fluctuation effect, i.e. phonon-phonon renormalization). To estimate the importance of the magnetic order as a source of deviation between theory and experiment, we have also calculated the phonon frequencies in the ferromagnetic (FM), antiferromagnetic (AFM) G-type and AFM-A type phases of PbVO3 (see Table S.6 in Supplemental Information). We found a strong spin-phonon coupling for some modes. For example, the E modes at 367 cm$^{-1}$ and at 274 cm$^{-1}$ are shifted by up to 28 cm$^{-1}$ depending on the magnetic ordering. This strong spin-phonon coupling does not explain the discrepancies observed between theory and experiment at high wavenumbers. For the purpose of this paper, it does not modify the mode assignment nor the following discussion of the eigenvectors.

Finally, we re-examine the tentative mode assignment proposed in Ref. [20] in the light of the current results. In this work, the assignment is based on a powder spectrum where, at variance with our experiments, $A_1$(LO) modes are allowed. The wavenumber of their $A_1$(3TO) agrees remarkably well with ours (876 vs. 874 cm$^{-1}$) and their value for the corresponding $A_1$(3LO) matches reasonably well the calculated LO-TO splitting value (66 vs. 78 cm$^{-1}$). They also observe two sharp peaks at 187 and 210 cm$^{-1}$ that match very well our values for the $A_1$(1TO) and $A_1$(1LO) respectively. The last $A_1$ mode is considerably weaker, but upon closer inspection, we do observe on their powder spectrum two unassigned peaks around 428 and 485 cm$^{-1}$ that match again very well our own $A_1$(2TO) wavenumber (429 cm$^{-1}$) and the calculated value for the LO-TO splitting (58 cm$^{-1}$). All the other peaks in the powder spectrum match the $B_1$ or E(TO) mode positions from our results. Those mode frequencies with their revised assignments are reported in Table 4.

Finally, as a side note, we observe that the amplitude of LO-TO splittings in PbVO3 remains quite moderate and correlate, via the Lyddane-Sachs-Teller relation, with relatively low values of the calculated dielectric permittivity ($\varepsilon_{11} = \varepsilon_{22} = 27$ and $\varepsilon_{33} = 13$).

3.5 Comparison between phonon modes in PbVO$_3$ and PbTiO$_3$.

PbTiO$_3$ and PbVO$_3$ are isostructural and have both the cubic perovskite as a parent high-symmetry structure; it is therefore tempting to establish links between their Raman modes. However, it is also natural to expect that the major differences in chemical bonding and strain result in significant mode mixing that make a one-to-one mapping hazardous. In order to explore the relevance of this comparison, we analyzed in more details the phonons of PbVO$_3$ and PbTiO$_3$. Specifically, in order to analyze the effect of the c/a ratio, we made the following computational experiment. We computed an artificial PbTiO$_3$ cell, named PTO-ST, whereby the volume and the c/a ratio were fixed to the experimental values of PbVO$_3$ and the internal atomic coordinates relaxed. Doing so, we obtained the relaxed atomic positions (Table S.7 in Supplemental Information), the born effective charges (Table S.8 in Supplemental Information), and hence very similar polarization values (117 µC/cm$^2$ in PbTiO$_3$ vs 132 µC/cm$^2$ in PbVO$_3$). In other words, PbTiO$_3$ strained artificially to a supertetragonal cell adopts a microscopic structure very close to that of PbVO$_3$.

We then computed the phonon frequencies in this artificial "supertetragonal" phase of PbTiO$_3$, which we report in Table 5. To highlight how the phonons of PTO-ST compare with those of PbVO$_3$, we report the overlap (i.e. projections of mode eigenvectors) between PbTiO$_3$ and PbVO$_3$ eigenvectors. We observe that all the PbTiO$_3$ eigenvectors have an almost perfect correspondence with those of PbVO$_3$ with overlap values above 0.87 in all cases. Most phonon frequencies show moderate differences with two remarkable exceptions. First, the high frequency $A_1$ mode shows a large discrepancy of 139 cm$^{-1}$. Second, the characters of the intermediate E modes are swapped and the overlaps show a small mode mixing. Even so, it is legitimate to consider that the overlap between phonon eigenvectors is very good and to establish a one-to-one correspondence between the modes in $PbTiO_3$ and $PbVO_3$. Overall, this indicates that the difference in chemistry between $PbTiO_3$ and $PbVO_3$ does not affect too much the phonon eigenvectors if they are calculated at the same cell parameters.

The correspondence between the experimental Raman signatures of $PbVO_3$ and $PbTiO_3$ is presented schematically in Figure 8. We report in black and red colors the two specific polarization configurations that highlight either the E or $A_1$ modes in both single crystals. The Raman spectrum of $PbTiO_3$ reported on the top was measured on a reference single crystal. The vertical bars and the dotted lines between them mark the band positions and the correspondence between Raman modes of each E, $A_1$ or $B_1$ symmetries as the c/a ratio increases from c/a=1.06 in $PbTiO_3$ to c/a=1.25 in $PbVO_3$.

A remarkable feature of the phonon spectrum of $PbVO_3$ as compared with $PbTiO_3$ is the sizeable splitting between the modes $B_1$ and E stemming from the silent $T_{2u}$ mode from the cubic phase. This splitting is in principle allowed in uniaxial crystals but is generally only significant for the $A_1$ and E modes [37], while the spectral separation between $B_1$ and E mode is usually so small that it cannot be resolved [51]. Here, we measured a separation of 50 cm$^{-1}$ between the $B_1$ and E modes at 319 and 269 cm$^{-1}$ respectively, in good agreement with the 45 cm$^{-1}$ calculated by DFT. Even larger $B_1$+E spectral separation of 208 cm$^{-1}$ were calculated from DFT simulations in $BiCoO_3$ compound [52] that display a larger tetragonal strain, c/a=1.267(1) [4], compared to $PbVO_3$. A sizeable splitting of the E and $B_1$ modes appears therefore as a characteristic feature of supertetragonal phases.

The comparison between the experimental Raman signatures of $PbVO_3$ and $PbTiO_3$ (see Fig. 8) shows that almost all phonons wavenumbers are higher in $PbVO_3$ as compared to $PbTiO_3$, in spite of the fact that the volume is larger. This is particularly true for the E(1TO) and the $A_1$(1TO) lowest wavenumber modes. These modes involve antisymmetric displacements either along the c-axis or in the x/y plane of the Pb atoms against the rigid $VO_5$ in $PbVO_3$ and can be considered as sensitive to the strength of the lead-oxygen bonds. The fact that we observe a shift toward higher wavenumbers as the volume increases indicates that the lead-oxygen force constant is increasing from $PbTiO_3$ to $PbVO_3$. Thus, our Raman measurements show some hardening of the Pb-O bond in $PbVO_3$. This observation agrees with recent first principle calculations of electron density in the $PbTi_{1-x}V_xO_3$ solid solutions that have shown that the vanadium substitution enhances the covalent interaction of the Pb−O bonds [53] which is expected to increase its bond strength.

3.6 Comparison with the modes of the cubic perovskite: tracking the ferroelectric soft mode

$PbTiO_3$ serves as a model for soft-mode driven ferroelectrics. As such, its soft optic mode and its lattice dynamics have been the subject of many works over decades in order to understand their details and their relation to the dielectric properties [54,55,56,57]. As it is well known, when calculated by first principles, cubic $PbTiO_3$ has one unstable $T_{1u}$ mode whose eigenvector also corresponds to the atomic displacements required to take cubic $PbTiO_3$ to the tetragonal phase [58]. This makes it a classical "soft-mode". It is however only one of the three $T_{1u}$ modes allowed so that hybridization with other $T_{1u}$ modes is in principle possible, either under the effect of temperature and anharmonicity, or by chemical substitution. The phase transition in $PbVO_3$ is experimentally very different in character, it occurs under pressure at 2.5 GPa as a strongly first-order transition with a very large volume drop of 10% [12]. It is not known at the moment whether phonon modes show significant softening experimentally, but the soft mode concept can be regardless used and discussed based on the phonon eigenvectors.

In our calculations with the HSE06 functional and the CRYSTAL code, we could not access to the simulation of phonons in the cubic phase of $PbVO_3$ due to the closure of the band gap and the disappearance of the magnetic order, so that we cannot compare the cubic phonons of $PbVO_3$ with those of its supertetragonal phase. However, as the phonon eigenvectors of supertetragonal $PbTiO_3$ are similar to those of $PbVO_3$, we can see how the phonon eigenvectors of ground state $PbVO_3$ project onto those of cubic $PbTiO_3$. We have calculated the phonons of the cubic phase of $PbTiO_3$ and in Table 6 we show the projections of the $PbVO_3$ ground state phonon eigenvectors against those of cubic $PbTiO_3$. The eigendisplacements are reported in Table S.9 in Supplemental Information.

The overlaps of the mode eigenvectors are best discussed by separating the modes according to their symmetries. We leave here aside the $B_1$ mode which is not polar and also constrained by symmetry to overlap perfectly with the $T_{2u}$ silent mode of the cubic phase. Considering first the overlap of the soft mode with the $A_1$ modes in the ground state of $PbTiO_3$, we observe and confirm that the soft mode does not correspond to the lowest $A_1$ mode but spread over all three of them, with a dominant contribution on the intermediate $A_1$ mode at 367 cm$^{-1}$. The same is true when considering the $PbVO_3$, with an even larger overlap with the highest $A_1$ mode at 910 cm$^{-1}$. The lowest $A_1$ mode of $PbTiO_3$ overlaps dominantly with the first stable polar mode of the cubic phase, and this trend is even more pronounced in supertetragonal $PbVO_3$.

The analysis of the E modes leads to similar conclusions, albeit with some additional complexity. First of all, the E mode from $T_{2u}$ clearly has to be considered since it becomes polar at the phase transition and mixes with the neighboring E mode, as described previously. As a consequence, when considering the overlaps in Table 6, one should bear in mind the swapping of the intermediate E modes characters described in the previous section. With these precautions however, it is clear that the trend observed for the $A_1$ modes is also reflected in the E mode eigenvectors. It therefore appears that there is a continuity in the lattice dynamics of $PbTiO_3$ and $PbVO_3$, whereby the soft phonon mode keeps shifting upwards, gradually transferring its character to modes of higher frequencies, while the displacements of the two other polar modes, in contrast, are shifted downwards and are dominantly reflected in the low frequency modes.

4. Conclusions

We have reported a combined theoretical and experimental study of the lattice dynamics of tetragonal $PbVO_3$ and carried out a detailed comparison with the isostructural compound $PbTiO_3$. All zone-center optical phonon modes in $PbVO_3$ have been identified by a combined polarized, angle-dependent Raman spectroscopy and first-principles density functional theory simulations. The comparison between the phonon eigenvectors shows that in spite of their differences in chemistry and strain, $PbVO_3$ and $PbTiO_3$ are remarkably similar, and that the soft mode concept can be carried over to $PbVO_3$ by considering the evolution of its projection onto the polar (and Raman active) modes. Differences in phonon frequencies, however, clearly reflect changes in bond strength that accompany modifications in chemical bonding. This, in particular, results in a lower value for the V-O vanadyl stretching mode as compared to other vanadium-based compounds. This work will be instrumental in studies of potential $Pb(Ti,V)O_3$ solid solutions and their characterization by Raman spectroscopy. Finally, we have also pointed out a strong dependence of some phonon frequencies upon magnetic order, i.e. a very strong spin-phonon coupling that will have to be investigated in future works, in relation to clarifying the magnetic ground state of $PbVO_3$.

Acknowledgements

PB and JAM acknowledge the French CNRS and C. Felix, Ch. Bouchard, O. Leynaud and A. Prat (Institut Néel) for their technical support. PB acknowledges C. Darie, C. Colin and P. Bordet for providing us the powder synthesized at high-temperature and high-pressure from which were found the crystals used in this study and V. Dmitriev for critical reading of the manuscript. JAM acknowledges Fundação para a Ciência e Tecnologia for the Sabbatical Grant SFRH/BSAB/142869/2018. AS and EB acknowledge the FRS-FNRS for funding. AS and EB also acknowledge the CECI supercomputer facilities funded by the F.R.S-FNRS (Grant No. 2.5020.1), the Tier-1 super- computer of the Fédération Wallonie-Bruxelles funded by the Walloon Region (Grant No. 1117545) and the OFFSPRING PRACE project (DECI resource BEM supercomputer based in Poland at Wrocław). The authors are grateful to J. Hlinka for providing a reference $PbTiO_3$ single crystal.

# Lattice dynamics and Raman spectrum of supertetragonal PbVO$_3$.

P. Bouvier[1], A. Sasani[2], E. Bousquet[2], M. Guennou[3], J. Agostinho Moreira[4]

[1] *Université Grenoble Alpes, Institut Néel CNRS, 25 Rue des Martyrs, 38042, Grenoble, France.*

[2] *Université de Liège, CESAM, QMAT, Physique Théorique des Matériaux, 19 Allée du 6 aout, B-4000, Sart Tilman, Belgium.*

[3] *Department of Physics and Materials Science, University of Luxembourg, 41 rue du Brill, L-4422 Belvaux, Luxembourg.*

[4] *IFIMUP – Instituto de Física de Materiais Avançados, Nanotecnologia e Fotónica, Departamento de Física e Astronomia, Faculdade de Ciências da Universidade do Porto. Rua do Campo Alegre s/n. 4169-007 Porto, Portugal.*

**Figures**

Figure 1. Sketch of the two reference frames chosen in this work. The laboratory reference frame (*X*,*Y*,*Z*) is fixed. The Y-axis is parallel to both the incident laser beam ($\vec{k}_i$) and the scattered signal ($\vec{k}_s$) The V and H polarization directions of the incident and scattered electric field are kept fixed along X and Z, respectively. The sample referential frame (a,b,c) oriented along the tetragonal crystallographic axes is rotated by $\theta$ over 180°.

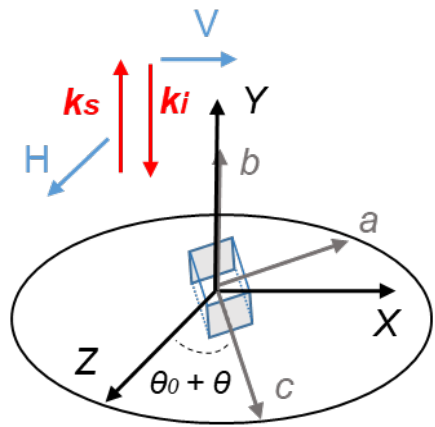

Figure 2. Representative Raman spectra of PbVO$_3$ recorded in the 100–1000 cm$^{-1}$ spectral range in the HV-scattering geometry at room conditions using the 514.5 nm and 488 nm laser lines as excitations. The wavenumbers (in cm$^{-1}$) and the symmetry of the eight observed first-order Raman bands are reported over each band.

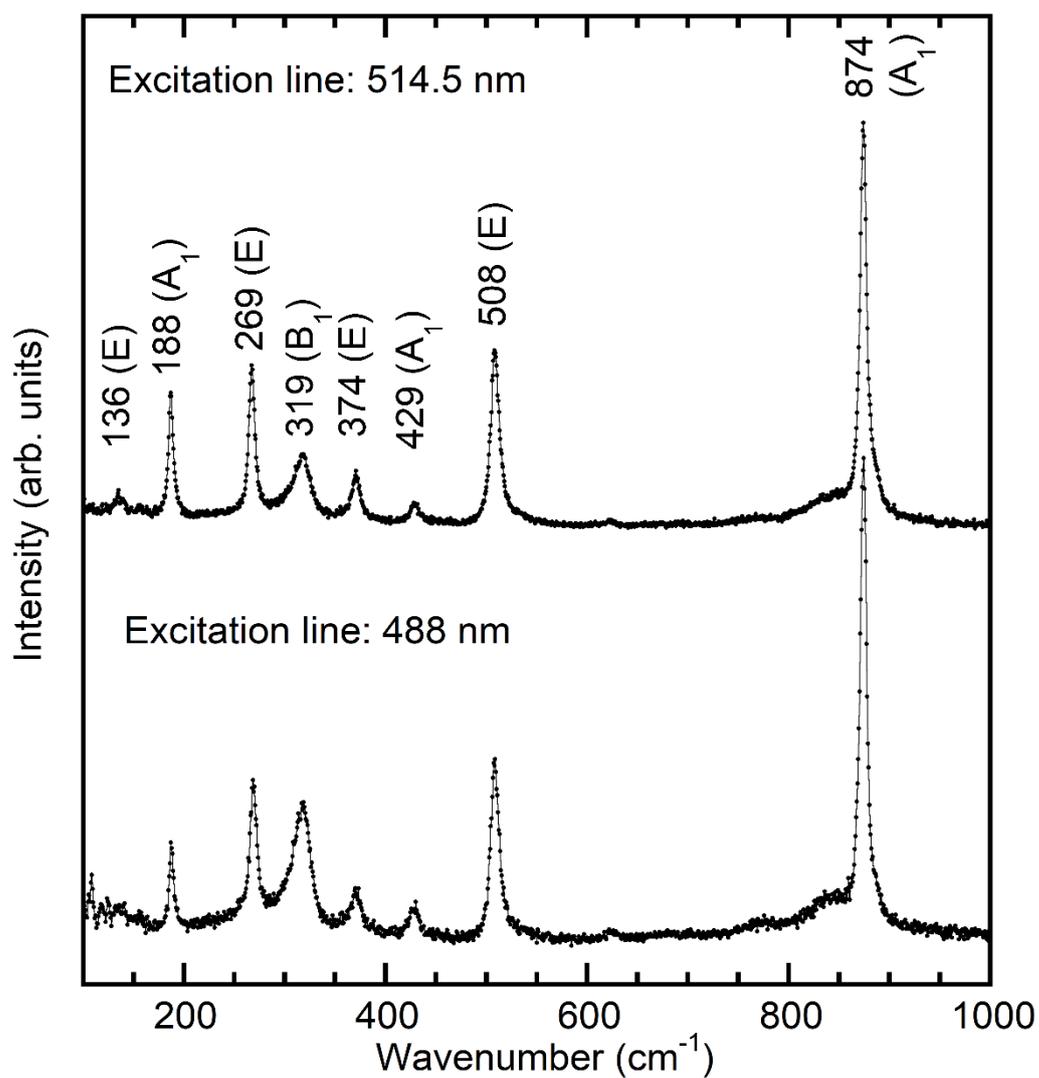

Figure 3. Raman spectra of PbVO$_3$ recorded, at room conditions, in the 100-1000 cm$^{-1}$ spectral range using the 514.5 nm line as excitation, for the (a) VV-, (b) HV-scattering geometries at some selected azimuthal angles. The angles are indicated on the side of each spectrum.

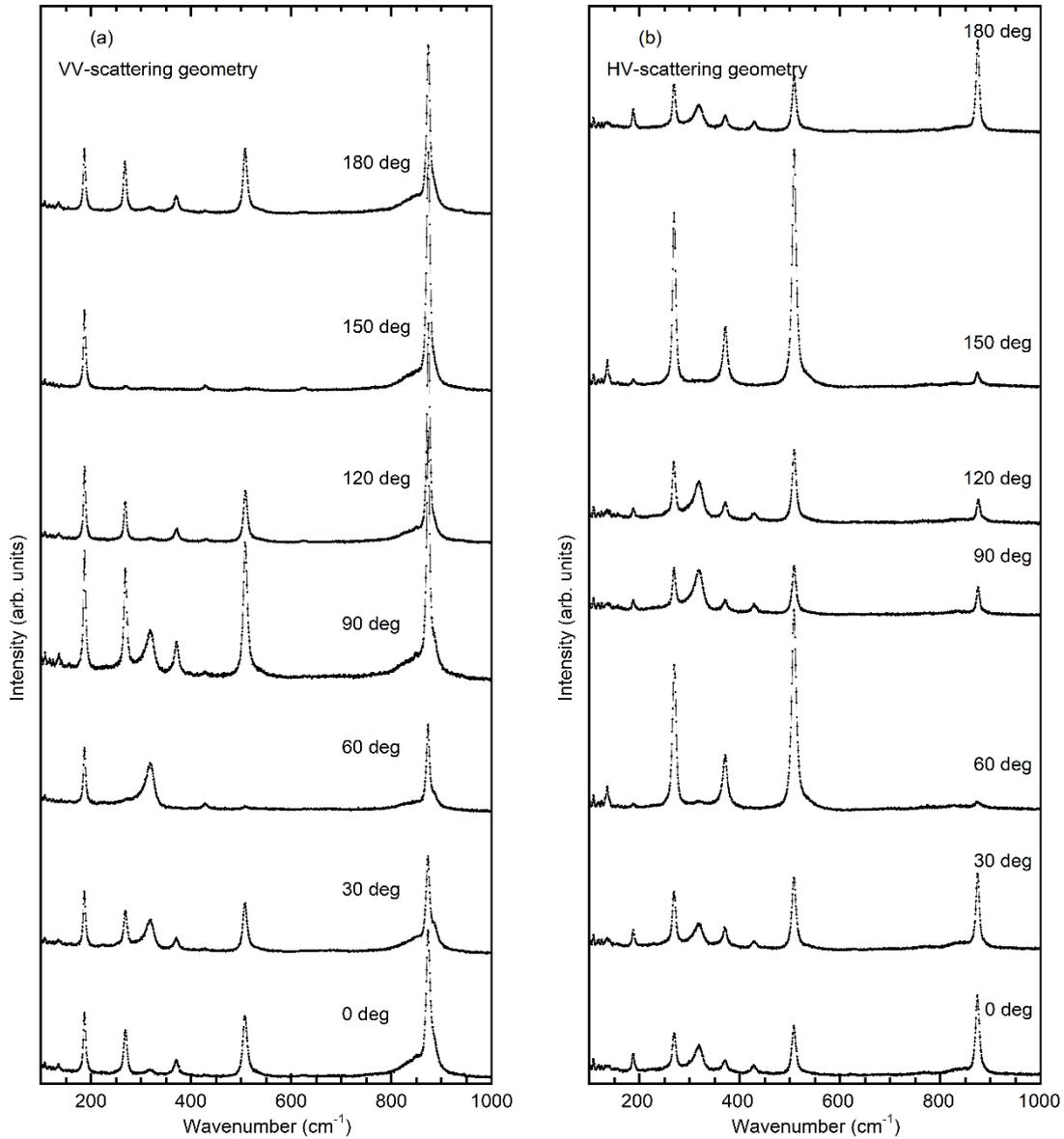

Figure 4. Polar plot of the intensity of the Raman bands centered at (a) 136, (b) 269, (c) 374, and (d) 508 cm$^{-1}$, measured in the VV and VH scattering geometries. The solid lines were determined through the best fit of Equations (14) and (15) to the experimental results, respectively.

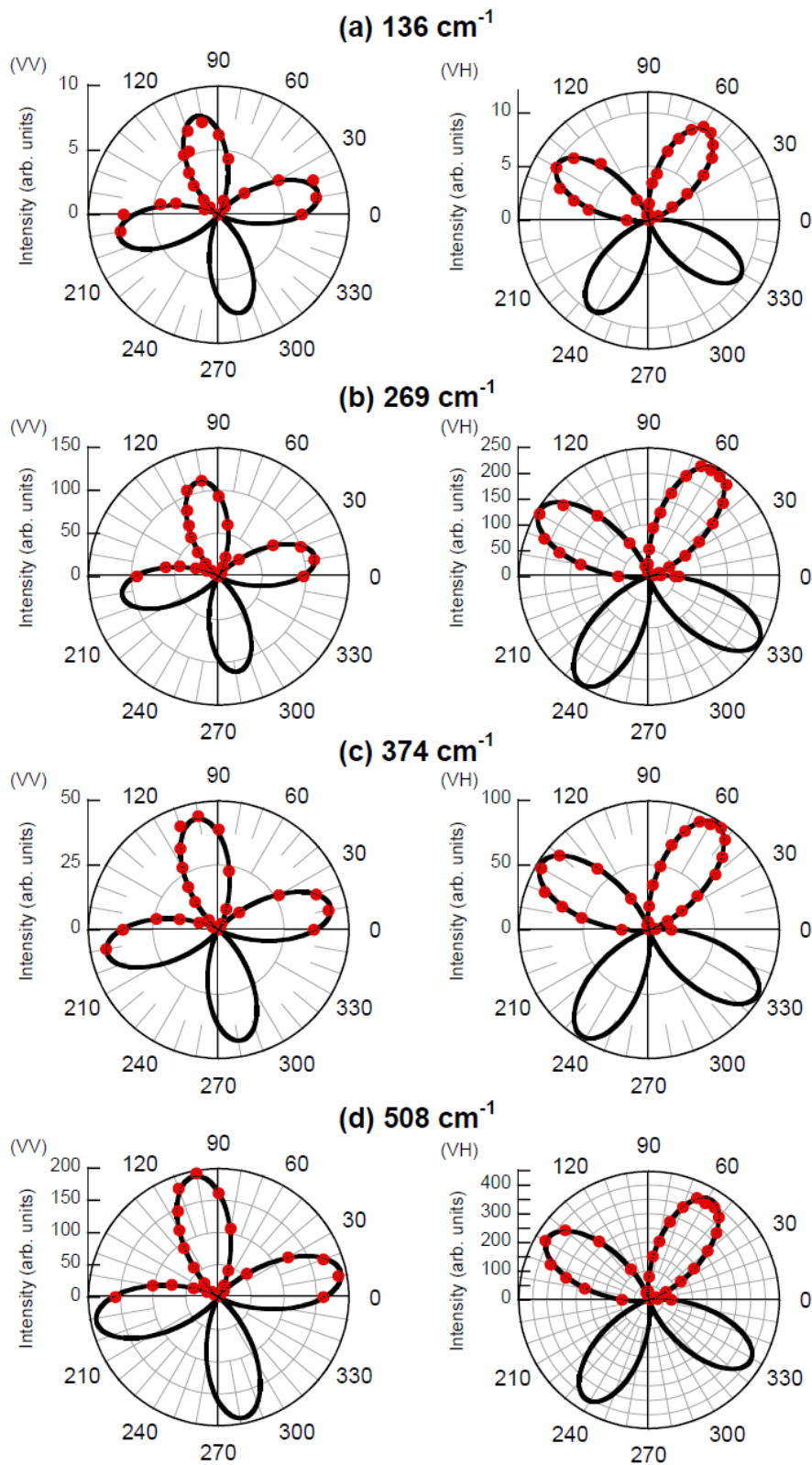

Figure 5. Polar plot of the intensity of the Raman bands centered at (a) 188, (b) 429, and (c) 874 cm$^{-1}$, measured in the VV and VH scattering geometries. The solid lines were determined through the best fit of Equations (8) and (9) to the experimental results, respectively.

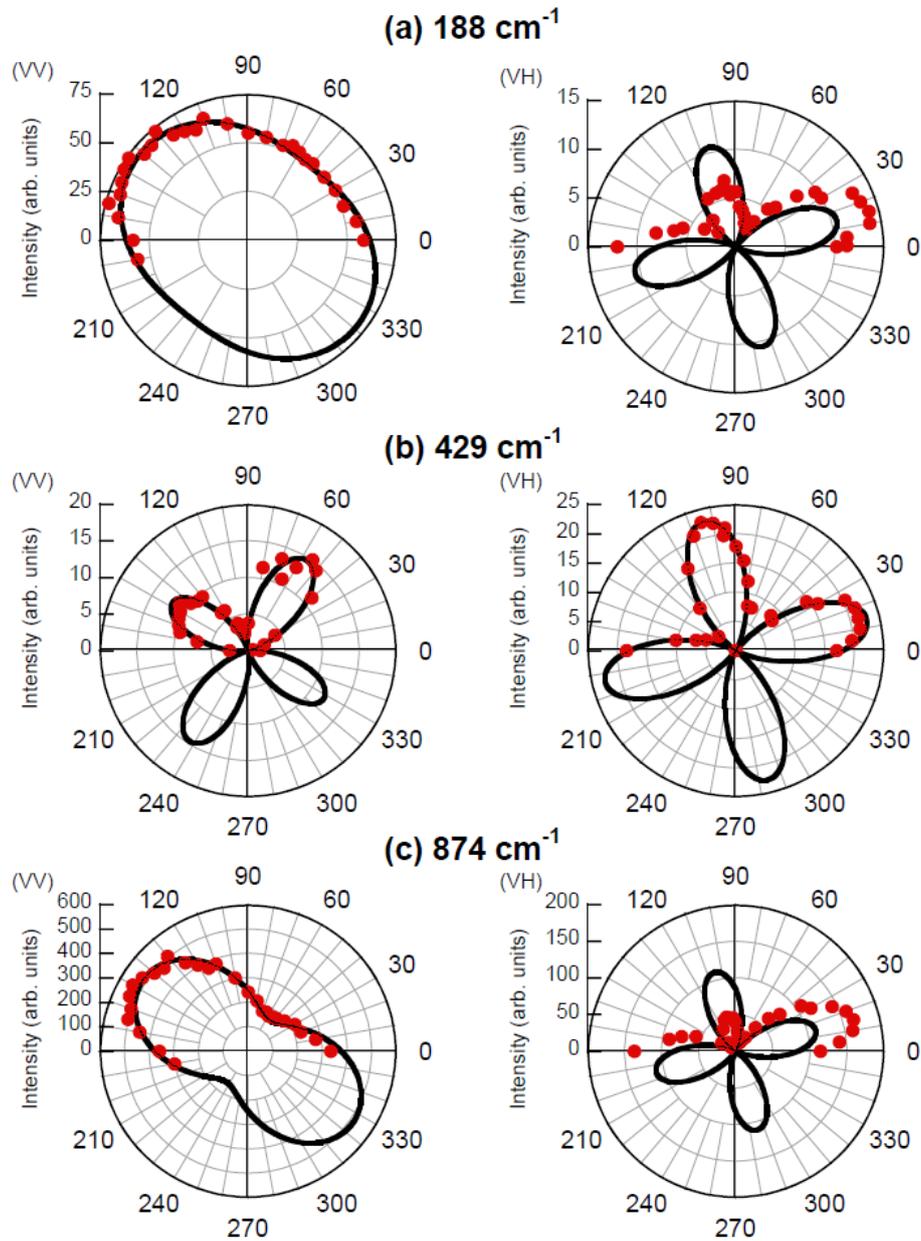

Figure 6. Polar plot of the intensity of the Raman band centered at 319 cm$^{-1}$, measured in the VV and VH scattering geometries. The solid lines were determined through the best fit of Equations (11) and (12) to the experimental results, respectively.

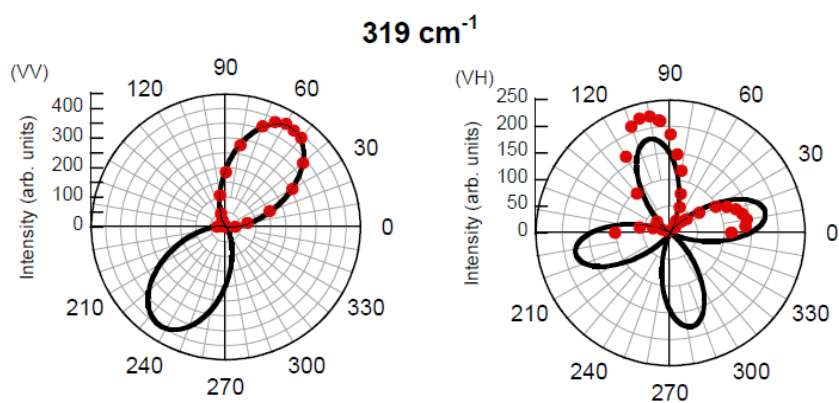

Figure 7: Schematic representation of the modes eigendisplacements and their associated wavenumbers calculated from our DFT simulations. The gray, gold and red balls corresponds to Pb, V and O atoms respectively. The arrows show the displacement of each atom and their different colors (blue vs brown) gives the sign of the atom displacements. The tetragonal *a*-, *b*- and *c*- crystallographic axes are reported on the left down side of each figure.

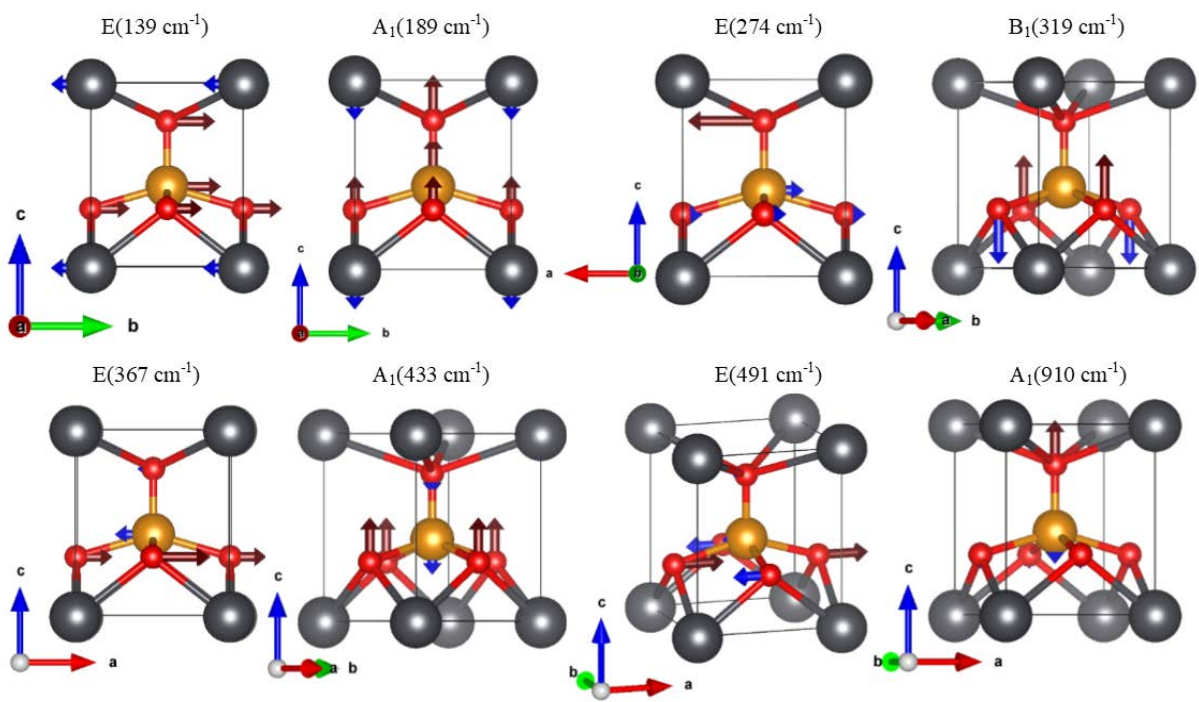

Figure 8: Sketch of the correspondence between the transversal optic (TO) Raman active phonons in PbTiO$_3$ (top) and in PbVO$_3$ (bottom) single crystals. The lines between both spectra highlight the hypothetical increase of the tetragonal ratio c/a from 1.06 to 1.25. The black and red colors correspond to two specific polarization configurations to highlight the E or A$_1$ modes, respectively.

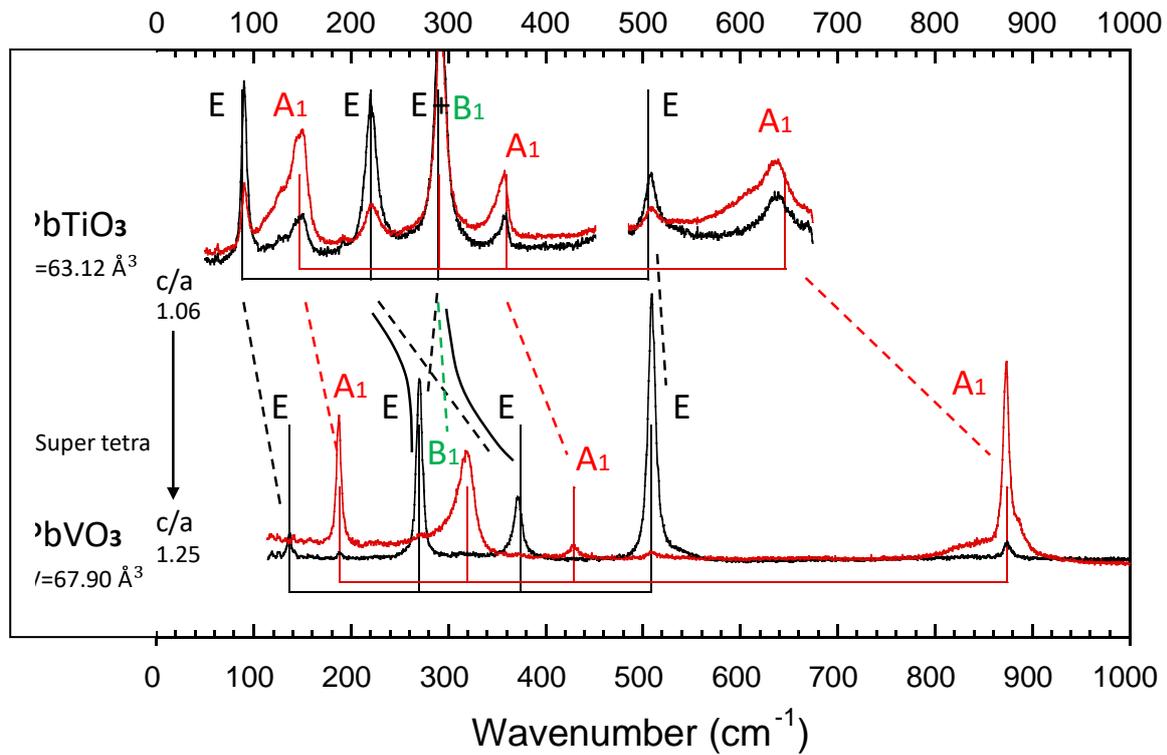

# Tables

Table 1. Occupied Wyckoff positions in the primitive cell of PbVO$_3$ and their contribution to the Raman-active modes. The Pb atoms are placed at the origin of the unit cell as in reference [10].

| Atom | Wyckoff position | A$_1$ | B$_1$ | E |
|---|---|---|---|---|
| Pb | 1a | 1 |  | 1 |
| V/O(1) | 1b | 1 |  | 1 |
| O(2) | 2c | 1 | 1 | 2 |

Table 2. Raman selection rules applicable to tetragonal PbVO$_3$ in relevant backscattering geometries. In each case, the symbol indicates which element of the Raman tensors contribute to the intensity of the Raman mode. The prime ' in $\alpha$' is to stress that the value of the Raman tensor element is different for the TO and LO modes.

|  |  | A$_1$(z) | B$_1$ | E(x) |  | E(y) |  | Sum of expected modes |
|---|---|---|---|---|---|---|---|---|
| z(xx)z | LO | $\alpha'^2$ | $\gamma^2$ | TO | - | TO | - | 3A$_1$(LO) + B$_1$ |
| z(xy)z |  | - | - | - | - | - | - | - |
| x(zz)x |  | $\beta^2$ | - | - | - |  | - | 3A$_1$(TO) |
| x(zy)x | TO | - | - | LO | - | TO | $\delta^2$ | 4E(TO) |
| x(yy)x |  | $\alpha^2$ | $\gamma^2$ | - | - |  | - | 3A$_1$(TO) + B$_1$ |

Table 3. Angular dependence of the intensity of the A$_1$, B$_1$ and E Raman bands in the different polarization geometries.

| Symmetry | Scattering geometry | Intensity versus $\theta$ | Eq. |
|---|---|---|---|
| A$_1$ | VV | $I(\theta) \propto |\alpha \sin^2(\theta + \theta_o) + \beta \cos^2(\theta + \theta_o)|^2$ | (8) |
|  | VH/HV | $I(\theta) \propto \left|\dfrac{(\beta - \alpha)}{2} \sin 2(\theta + \theta_o)\right|^2$ | (9) |
|  | HH | $I(\theta) \propto |\alpha \cos^2(\theta + \theta_o) + \beta \sin^2(\theta + \theta_o)|^2$ | (10) |
| B$_1$ | VV | $I(\theta) \propto |\gamma \sin^2(\theta + \theta_o)|^2$ | (11) |
|  | VH/HV | $I(\theta) \propto \left|\dfrac{\gamma}{2} \sin 2(\theta + \theta_o)\right|^2$ | (12) |
|  | HH | $I(\theta) \propto |\gamma \cos^2(\theta + \theta_o)|^2$ | (13) |
| E | VV | $I(\theta) \propto |\delta \sin 2(\theta + \theta_o)|^2$ | (14) |
|  | VH/HV | $I(\theta) \propto |\delta \cos 2(\theta + \theta_o)|^2$ | (15) |
|  | HH | $I(\theta) \propto |\delta \sin 2(\theta + \theta_o)|^2$ | (16) |

Table 4. Experimental and theoretical positions (for the ground state C-type magnetic ordering), and corresponding symmetry assignment of the first-order Raman bands of PbVO3 in the tetragonal P4mm structure. The values in brackets are the LO mode frequencies including the non-analytical correction to the modes frequencies. Theoretical and experimental results from Refs. 15 and 20 respectively are added for completeness. For Ref. 20, the wavenumbers are from the paper but the symmetry assignment is revised in the light of the present work. The star* indicates weaker modes that are visible on the spectrum but not assigned in the paper itself.

| Experimental (this study) | | DFT (HSE06, this study) | | DFT (LAPW) ref. [15] | Experimental ref. [20] |
|---|---|---|---|---|---|
| Wavenumber TO (cm$^{-1}$) | Symmetry | Wavenumber TO (LO) (cm$^{-1}$) | Symmetry | Wavenumber (cm$^{-1}$) | Wavenumber TO (LO) (cm$^{-1}$) |
| 136 | E(x,y) (1TO) | 139 (151) | E | | 140* |
| 188 | $A_1$(z) (1TO) | 189 (212) | $A_1$ | 190 | 187 (210) |
| 269 | E(x,y) (TO) | 274 (278) | E | | 263 |
| 319 | $B_1$ | 319 | $B_1$ | | 316 |
| 374 | E(x,y) (2TO) | 367 (394) | E | | 370 |
| 429 | $A_1$(z) (2TO) | 433 (491) | $A_1$ | 408 | 428* (485*) |
| 508 | E(x,y) (3TO) | 491 (502) | E | | 506 |
| 874 | $A_1$(z) (3TO) | 910 (988) | $A_1$ | 838 | 876 (942) |

Table 5. Projection of the phonon eigenvectors calculated for the ground state of PbVO$_3$ (PVO-GS) against the phonon eigenvectors of an artificial supertetragonal PbTiO$_3$ calculated at a c/a ratio equal to 1.25 (PTO-ST). A perfect match gives an overlap of 1. The largest overlaps are highlighted by the color scale.

| PTO ST \ PVO GS | | E modes | | | | A$_1$ modes | | | B$_1$ |
|---|---|---|---|---|---|---|---|---|---|
| | | 139 | 274 | 367 | 491 | 189 | 433 | 910 | 319 |
| E modes | 108 | 0.99 | 0.00 | 0.00 | 0.00 | | | | |
| | 271 | 0.00 | 0.13 | 0.88 | 0.00 | | | | |
| | 277 | 0.00 | 0.87 | 0.11 | 0.00 | | | | |
| | 538 | 0.00 | 0.00 | 0.00 | 0.99 | | | | |
| A$_1$ modes | 178 | | | | | 0.99 | 0.02 | 0.00 | |
| | 423 | | | | | 0.01 | 0.98 | 0.00 | |
| | 771 | | | | | 0.00 | 0.00 | 1.00 | |
| B$_1$ | 282 | | | | | | | | 1 |

Table 6: Projection of the phonon eigenvectors of the PbVO$_3$ and PbTiO$_3$ ground states (PVO GS and PTO GS respectively) against the ones of the cubic phase of PbTiO$_3$. By symmetry, the B$_1$ mode has a perfect overlap with the cubic T$_{2u}$ mode and is omitted here. The largest overlaps are highlighted by the color scale. The stars* are to stress the swapping of the two E modes as discussed in the text and shown in Table 5. Wavenumbers in cm$^{-1}$.

| | | | | A$_1$ modes | | | E modes | | | |
|---|---|---|---|---|---|---|---|---|---|---|
| | | | PTO | 152 | 367 | 678 | 104 | 213 | 291 | 515 |
| | | | PVO | 189 | 433 | 910 | 139 | 367* | 274* | 491 |
| Modes of cubic PTO | T$_{1u}$(1) | -161 | on PTO | 0.29 | 0.53 | 0.18 | 0.36 | 0.62 | 0.00 | 0.01 |
| | T$_{1u}$(1) | -161 | on PVO | 0.16 | 0.55 | 0.30 | 0.15 | 0.59 | 0.24 | 0.02 |
| | T$_{1u}$(2) | 117 | on PTO | 0.71 | 0.19 | 0.10 | 0.63 | 0.34 | 0.02 | 0.00 |
| | T$_{1u}$(2) | 117 | on PVO | 0.81 | 0.03 | 0.13 | 0.83 | 0.06 | 0.07 | 0.02 |
| | T$_{2u}$ | 246 | on PTO | | | | 0.00 | 0.03 | 0.93 | 0.03 |
| | T$_{2u}$ | 246 | on PVO | | | | 0.00 | 0.35 | 0.53 | 0.12 |
| | T$_{1u}$(3) | 501 | on PTO | 0.00 | 0.27 | 0.72 | 0.00 | 0.01 | 0.04 | 0.95 |
| | T$_{1u}$(3) | 501 | on PVO | 0.03 | 0.42 | 0.56 | 0.00 | 0.00 | 0.16 | 0.84 |